
\magnification=1200
\pageno=0
\nopagenumbers

\hskip 9.5cm \vbox{\hbox{DFTT 24/91}\hbox{\it revised
version}\hbox{\rm
November 1991}}
\vskip 1cm
\centerline{\bf COMMENTS ON SUPERSYMMETRIC }
\centerline{\bf VECTOR AND MATRIX MODELS}
\vskip 1.5cm
\centerline { A. D'Adda$^*$}
\vskip .5cm
\centerline{\sl Istituto Nazionale di Fisica Nucleare, Sezione di Torino}
\centerline{\sl  Dipartimento di Fisica
Teorica dell'Universita' di Torino}
\centerline{\sl via P.Giuria 1, I-10125 Turin, Italy}
\vskip .8cm

\vskip 1.5cm
\centerline{\bf Abstract}
\vskip 0.5cm
\noindent
Some results in random matrices are generalized to supermatrices, in
particular supermatrix integration is reduced to an integration over the
eigenvalues and the resulting volume element is shown  to be equivalent
 to a one dimensional Coulomb gas of both positive
  and negative charges.It is shown that,for polynomial potentials,
  after removing the instability due to the annihilation of opposite
 charges, supermatrix models are indistinguishable from ordinary matrix
 models, in agreement with a recent result by Alvarez-Gaum\`e and
 Ma\~nes. It is pointed out however that this may not be true for
 more general potentials such as for instance the supersymmetric
generalization of the Penner model.

\vskip 2.2cm
\hrule
\vskip .8cm
\hbox{\vbox{\hbox{$^*${\it email address:}}\hbox{}}
 \vbox{\hbox{ Decnet=(39163::DADDA)}
\hbox{ Bitnet=(DADDA@TORINO.INFN.IT)}}}

\vfill
\eject
\footline={\hss\tenrm\folio\hss}
\vskip 0.5 cm
One would na\"\i vely expect that  a model based on supermatrices is a
 generalization of matrix
 models [1] suitable to
describe a discretized two dimensional supergravity coupled with
superconformal matter. In fact it contains
both fermionic and bosonic degrees of freedom related by a local ( in a
zero dimensional sense) supersymmetry.

However in matrix models
the matrix elements are sort of preons whose interpretation in terms of
two dimensional space time is far from being obvious. So the
supersymmetry in supermatrix models may easily not have anything to do
with space-time supersymmetry.

We show in this note that this is indeed the case,and that, in
 supermatrix models with an action $S$ polynomial in the supermatrix
 $\Lambda$, the fermionic degrees of freedom simply cancel an equal
number of bosonic degrees of freedom, so that in the end one is left
with an ordinary matrix model with a reduced  (if any) number of
entries. This confirms the results obtained in other way in
 ref. [2]\footnote{${}^{*}$}{Most results contained in this paper, and
in particular eq.(13) have been obtained prior to our knowledge of ref.
[2]}.We point out however that with non polynomial actions such
cancellation between bosonic and fermionic degrees of freedom does not
necessarily occur, leading possibly to models that are not equivalent
ordinary matrix models.

We show as well that, in the same way as ordinary matrix models are
equivalent to a Coulomb gas of equal charges, whose positions are
 given by
the eigenvalues of the matrix, supersymmetric matrix models can be
described in terms of a Coulomb gas of both positive and negative
charges whose positions correspond to the eigenvalues of the
supermatrix.
The sign of the charge depends upon which bosonic submatrix the
corresponding eigenvalue belongs to.
The annihilation of a couple of opposite charges corresponds in this
picture to the cancellation of bosonic and fermionic degrees of freedom.

Before considering the relatively more complicated case of matrix models
let us study the supersymmetric generalization of the vector models
analyzed in [3], where supersymmetry leads to a rather straightforward
cancellation between bosons and fermions.
Consider a vector $\vec v$ consisting of $n$ bosonic and $m$ fermionic
components:
$$\vec v =\pmatrix{x_1\cr .\cr .\cr x_n\cr \psi_1\cr .\cr .\cr \psi_m}$$
and a partition function given by
$$Z= \int d^nx d^m\psi e^{-S[\vec v^2]}\eqno(1)$$
where $ \vec v^2 $ is the $OSp(n,m)$ invariant scalar product:
$$\vec v^2=\sum_{i=1}^n x_i^2+\sum_{\alpha=1}^m \psi_{\alpha} C^
{\alpha\beta}\psi_{\beta}$$
and $C^{\alpha\beta}$ is a symplectic matrix.

One can compute $Z$ by gauge fixing the $OSp(n,m)$ symmetry.\ With the
gauge fixing conditions
$$x_2=x_3=\cdots =x_n=\psi_1=\psi_2=\cdots =\psi_m=0$$
the Faddeev-Popov determinant is $x_1^{n-m-1}$ and the partition
 function becomes
$$Z=\int dx_1 x_1^{n-m-1} e^{-S(x_1^2)}\eqno(2)$$

Due to the cancellation between fermions and bosons the partition
function (2) coincides with the one that one would obtain from a purely
bosonic $n-m$ dimensional vector model.

Let us consider now a supermatrix model defined by the partition function
$$Z=\int \prod_{A,B=1}^{n+m} d\Lambda_{\ B}^A e^{-S[\Lambda]}\eqno(3)$$

where $ \Lambda$ is a supermatrix and it has the following block
 structure:
$$\left\{\Lambda_{\ B}^A\right\}\ =\ \left\{\matrix{M_{\ b}^a&\psi_{\
\beta}^a\cr\overline{\psi}_{\ b}^{\alpha}&N_{\ \beta}^{\alpha}}
\right\}\quad {\rm where}\quad \matrix{a,b=1,2,\cdots,m \cr\alpha,\beta=1
,2,\cdots,n}\eqno(4)$$
where $M_{\ b}^a$  and$ N_{\ \beta}^{\alpha}$  are even Grassmann
variables while $\psi_{\ \beta}^a$ and $\overline{\psi}_{\ b}^{\alpha}$
 are Grassmann odd.In what follows we shall consider the case where
$\Lambda$ is a hermitian supermatrix and the action $S[\Lambda]$ is
invariant under the infinitesimal $U(m|n)$ transformations
$$\delta\Lambda=[\epsilon,\Lambda]\eqno(5)$$
where $\epsilon$ is an antihermitian supermatrix and hence a generator
of the supergroup $U(m|n)$.
 The action $S$ can be a function only of the $U(m|n)$ invariants:
$$\eqalign{str(\Lambda^k) \ &=\ \sum_A (-1)^{\tau_A} \ (\Lambda^k)_{\ A}
^A \ =\
\sum_a(\Lambda^k)
_{\ a}^a-\sum_{\alpha}(\Lambda^k)_{\ \alpha}^
{\alpha}\cr &=\ \sum_a \lambda_a^{\ k}-\sum_{\alpha}\mu_{\alpha}^{\ k}
\cr}\eqno(6)$$
 where $\lambda_a$ and $\mu_{\alpha}$ are the eigenvalues of $\Lambda$
and also form a complete set of independent $U(m|n)$ invariants.

The partition function $Z$ however, as given by the r.h.s. of eq.(3), is
not well defined as the integral either gives identically zero or is in
fact divergent.
 This can be seen directly in
the $m=n=1$ case where $\Lambda$ is given by
$$\Lambda\ =\ \pmatrix{\tilde {\lambda} & \psi \cr \bar {\psi} & \tilde
{\mu} \cr}\eqno(7)$$
and the eigenvalues of $\Lambda$ are:
$$\lambda\ =\ \tilde {\lambda} + {\psi \bar {\psi} \over \tilde{\lambda}
- \tilde{\mu}}\ \ \ \ \ \ \ \ \ \mu\ =\ \tilde{\mu} + {\psi \bar{\psi}
\over \tilde{\lambda} - \tilde{\mu}}\eqno(8)$$
After defining $\xi_+=\tilde{\lambda}+\tilde{\mu}$ and $\xi_-=
\tilde{\lambda}-\tilde{\mu}$ the partition function can be written as:
$$\eqalign{Z\ & =\ \int d\xi_+d\xi_-d\psi d\bar{\psi}e^{-S(\xi_++{2\psi
 \bar{\psi}
\over \xi_-},\xi_-)}\cr &=-2\int d\xi_+d\xi_- {1 \over \xi_-} {\partial
S \over \partial \xi_+} e^{-S(\xi_+,\xi_-)}\cr}\eqno(9)$$
The integral at the right hand side of eq.(9) can diverge at $\xi_- = 0$
or, if $S$ vanishes at $\xi_- =0$, at infinity in the $(\xi_+,\xi_-)$
plane. This is the case for instance of any action of the type
 $$ S=\sum_{k=2}^l g_k str(\Lambda^k)\eqno(10)$$
 On the other hand, if none of these
divergences occur, $Z$ vanishes as the integrand is a  total derivative
in $\xi_+$. This latter property is a consequence of supersymmetry and
it is completely general: the integral at the r.h.s. of eq.(3) involves
the integration over $2nm$ fermionic variables while $S(\Lambda)$
 is a function only of the eigenvalues of $\Lambda$.
The divergence of the integral in eq.(3) for any action of the type
 given in eq.(10) is also a  property valid for the general
 $U(n|m)$ model; in this
respect supermatrix models are very different from ordinary matrix
models where the integral defining the partition function is well
defined at least for actions analogue to (10) with even $l$ and suitable
coupling constants. Besides,
in matrix models the perturbative expansion of $Z$ in powers of $g_k$
($k>2$) can always,at least formally, be defined as its coefficients are
given in terms of convergent integrals, while in supermatrix models such
coefficients are divergent as already the quadratic term in (10) in not
positive definite.

In order to define $Z$ in a meaningful way we first
 reduce the integral at the r.h.s. of (3) to an integral over gauge
invariant quantities, i.e. the eigenvalues of $\Lambda$. In doing that
it is essential to preserve the $U(n|m)$ invariance , so we fix the
gauge in which $\Lambda$ is diagonal and compute the corresponding
Faddeev-Popov superdeterminant.
In the chosen gauge the variation of $\Lambda$ under an infinitesimal
supergauge transformation is given by:
$$\delta\Lambda_{\ B}^A=\epsilon_{\ B}^A(\lambda_B-\lambda_A)\eqno(11)$$
where $\lambda_A$ are the eigenvalues of $\Lambda$. The Faddeev-Popov
 superdeterminant is then given by:
$$sdet\left\{{\delta\Lambda_{\ B}^A\over \delta\epsilon_{\ D}^C}\right\}=
\prod_{A> B}(\lambda_B-\lambda_A)^{2\times(-1)^{\tau_A+\tau_B}}=
{\prod_{a> b}
(\lambda_a-\lambda_b)^2 \ \prod_{\alpha> \beta}(\mu_{\alpha}-\mu_{\beta}
)^2\over \prod_{a,\alpha}(\lambda_a-\mu_{\alpha})^2} \eqno(12)$$
where $\{\lambda_A\}\equiv \{\lambda_a,\mu_{\alpha}\}$.

Eq. (12) is the supermatrix generalization of the volume element found
 by Mehta for ordinary hermitian matrices [4], and it reduces to it for
 $m=0$.
It follows that after fixing the gauge the partition function (3) for
 hermitian supermatrices becomes:
$$Z(n,m)=\int\ \prod_1^m d\lambda_a\prod_1^nd\mu_{\alpha}\ {\prod_{a> b}
 (\lambda_a-\lambda_b)^2\ \prod_{\alpha> \beta}(\mu_{\alpha}-\mu_
{\beta})^2\over\prod_{a,\alpha}(\lambda_a-\mu_{\alpha})^2}\ e^{-S[
\lambda_a , \mu_{\alpha} ]}\eqno(13)$$

By taking the volume element to the exponent one sees that the partition
function (13) describes a one-dimensional Coulomb gas of opposite charges
whose positions are given by the two sets of eigenvalues $\lambda_a$ and
$\mu_{\alpha}$ . The original action $S$ acts as an external
field.
The integral in (13) has two kind of divergences: one, corresponding to
the vanishing of the denominators in the integral, denotes the
instability of the system towards annihilation of opposite charges, the
other, which occurs for instance if $S$ is given by (10), is the result
of the instability originated by the external field that attracts
 the appropriate charges  to $\pm \infty$.
  The first type of divergences can be regularized by  restricting
the integration volume in eq. (13) to the region defined by the
inequalities $|\lambda_a-\mu_{\alpha}|<\epsilon$. In the limit $ \epsilon
\rightarrow 0$ the integral is dominated by the configurations where the
maximum number of denominators vanish. More precisely we find :
$$Z(n,m)\propto\epsilon^{-m} Z(n-m,0)+O(\epsilon^{-m+1})\eqno(14)$$
where we assumed $n\geq m$ and $Z(n-m,0)$ coincides with the partition
function of a model of  ordinary $(n-m)\times(n-m)$ matrices.
\ Consequently, the correlation functions of the $U(n,m)$ invariant
operators $O_r = str\Lambda^r$ coincide ,\ in the $\epsilon
 \rightarrow 0$ limit, with the corresponding correlation functions of
 an ordinary matrix model.\
The divergences that arise from the potential not being bounded can be
eliminated by replacing for instance the action (10) with
$$ S(\lambda_a,\mu_{\alpha}) = \sum_{k=2}^l g_k \Bigl[  \sum_{a=1}^n
 \lambda_a^k -\sum_{\alpha =1}^m \mu_{\alpha}^k \Bigr]
 + \eta \Bigl[\sum_{a=1}^n \lambda_a^{2p} +
 \sum_{\alpha=1}^m \mu_{\alpha}^{2p} \Bigr]\eqno(15)$$
where $\eta >0$ and $2p > l$. In the limit $\eta \rightarrow 0$ and
, for instance, $l$ even the integral is dominated by the
 configurations where one set of variables, let's say the $\lambda$'s,
are very large in absolute value (order $\eta^{-{1 \over 2p}}$).

Clearly  the two limits $\epsilon \rightarrow 0$ and
$\eta \rightarrow 0$ do not commute, and the order of the two limits has
to be given to define the theory. However the theory defined by  first
 taking the limit $\epsilon \rightarrow 0$ is the one that closely
resembles ordinary matrix models. In fact if one performs a perturbative
expansion of the partition function (3) in terms of Feynman diagrams, as
in ref. [2], the only divergent quantity one encounters is the
propagator. This needs to be regularized by  introducing a cutoff $\eta$
as in (15), or by any other equivalent regularization procedure (in ref.
[2] for instance a Wick rotation in the $\mu$'s is performed ).
 Such cutoff is kept to the end of the calculation. On the other hand by
performing the integral (3) over the matrix elements without fixing the
gauge one takes into account
automatically that configurations where couples of $\lambda$ and $\mu$
eigenvalues coincide have an infinite weight compared to the other
configurations, and the $\epsilon \rightarrow 0$ limit is automatically
done.

 Our analysis confirms the results of ref. [2] that
supermatrix models are equivalent to the ones based on ordinary
 matrices; however this conclusion should be restricted to models
 where the action, as in eq (10), is given by a linear combination of
 terms of the form $str\Lambda^k$. For more complicated potentials
 supermatrix models may differ substantially from ordinary matrix
 models. The supersymmetric extension of  Penner-type models [5] could
 be an interesting
example of this situation. Consider for example a supermatrix model
 coupled with $r$
complex supervectors .As a result of the integration on the supervectors
one gets in the integral at the r.h.s. of eq.(3) an extra factor $sdet(
\Lambda^r)$ which enhances the configurations with large $\lambda$'s
 and small  $\mu$'s . This effect can break the pairing of eigenvalues
due to supersymmetry and prevent a complete cancellation between bosonic
and fermionic excitations.

As a final remark it is worth noticing that, as in ordinary
 matrix models,
different choices of $\Lambda$ lead to different models:for instance
if $\Lambda$ has the form:
$$\Lambda_B^A=g^{AC}\Lambda_{CB}\eqno(16)$$ with
$$\Lambda_{CB}=(-1)^{\tau_C\tau_B}\Lambda_{BC}\eqno(17)$$
and
$$\left\{g^{AB}\right \}=\pmatrix {\delta^{ab}&0\cr 0&C^
{\alpha\beta}}\eqno(18)$$
with $C^{\alpha\beta}$ a symplectic metric,
then  the action is invariant under transformation (5) with $\epsilon$
 belonging to the algebra of the
orthosymplectic group and the corresponding Faddeev-Popov
 superdeterminant is given by:
$${\prod_{a> b}
|\lambda_a-\lambda_b| \ \prod_{\alpha >\beta}(\mu_{\alpha}-\mu_{\beta})^4
\over \prod_{a,\alpha}(\lambda_a-\mu_{\alpha})^2}\eqno(19)$$
In the last equation the greek indices run from $1$ to $m/2$ to take
 into account the fact that
 each eigenvalue $\mu_{\alpha}$ stands for two coincident eigenvalues
 of the symplectic
part of  $\Lambda$.
\vfill
\eject

\centerline{\bf References}
\vskip .5cm
\item{1.} For a review on matrix models see for instance:

A. Bilal,\it 2D Gravity from matrix models\rm,Cern-TH 5867/90

and references therein.
\vskip 0.2cm
\item{2.} L. Alvarez-Gaum\'e and J.L.Ma\~nes,\it Supermatrix Models\rm,
 Cern-TH.6067/91.
  \vskip 0.2cm
\item{3.} P. Di Vecchia,M. Kato and N. Ohta,\it Double scaling limit in
$O(n)$ vector models\rm , Nordita preprint 90/59 P.

S.Nishigaki and T. Yoneya, Tokyo preprint, UT-Komaba 90-16 (1990).
\vskip 0.2cm
\item{4.} M.L. Mehta,\it Random Matrices\rm,Academic Press 1967.
\item{5.} R.C. Penner,\it Bull. Am. Math. Soc.\rm 15 (1986) 73.

J. Distler and C. Vafa, "A Critical Matrix Model at $c=1$", Harward
preprint HUTP-90/A062, Oct. 1990.
\end